\documentclass[12pt]{iopart}
\usepackage{isotope}
\usepackage{multirow}
\usepackage{graphicx}
\usepackage{color}

\newcommand{\ket}[1]{\left| #1 \right\rangle}

\newcommand{\rket}[1]{\left\| #1 \right\rangle}
\newcommand{\rbra}[1]{\left\langle #1 \right\|}

\begin{document}

\title[Laser-induced electronic bridge with$~^{\rm{229m}}\rm{Th}\rightarrow ~^{\rm{229g}}\rm{Th}$ nuclear transition]{Laser-induced electronic bridge for characterization of the $\isotope[229\rm{m}]{Th}\rightarrow\isotope[229\rm{g}]{Th}$ nuclear transition with a tunable optical laser}

\author{Pavlo V. Bilous$^1$, Ekkehard Peik$^2$, Adriana P\'alffy$^1$}
\address{$^1$ Max-Planck-Institut f\"ur Kernphysik, Saupfercheckweg 1, D-69117 Heidelberg, Germany}
\address{$^2$ Physikalisch-Technische Bundesanstalt, Bundesallee 100
D-38116 Braunschweig, Germany}
\ead{Palffy@mpi-hd.mpg.de}

\begin{abstract}
An alternative method to determine the excitation energy of the $\isotope[229\rm{m}]{Th}$ isomer via the laser-induced electronic bridge is investigated theoretically. 
In the presence of an optical or ultra-violet laser at energies that fulfill a two-photon resonance condition, the excited nuclear state can decay by transfering its energy to the electronic shell.  A bound  electron is then promoted to an excited state by absorption of a laser photon and simultaneous de-excitation of the nucleus. We present calculated rates for the laser-induced 
electronic bridge process and discuss the experimental requirements for the corresponding setup. Our results show that depending on the actual value of the nuclear transition energy, the rate can be very high, with an enhancement factor compared to the radiative nuclear decay of up to $10^8$. 
\end{abstract}

%
%
%
\maketitle
%
%

\section{Introduction}

The $\isotope[229]{Th}$ isotope has been subject of increased attraction of the scientific community due to its  unique long lived excited state $\isotope[229m]{Th}$ at an energy lying in the VUV range~\cite{Wense_Nature_2016, Beck_78eV_2007, Beck_78eV_2007_corrected}. Such long-lived excited states are termed in nuclear physics  {\it isomers}.
The $\isotope[229\rm{m}]{Th}\rightarrow\isotope[229–\rm{g}]{Th}$  transition to the ground state has an extremely small ratio of the radiative width to the energy, estimated at~$10^{-19}$. This opens the possibility for exciting applications such as a very precise nuclear frequency standard~\cite{Peik_Clock_2003,Campbell_Clock_2012,Peik_Clock_2015} or a nuclear laser~\cite{Tkalya_NuclLaser_2011}. The extremely low energy may render nuclear spectroscopy and even coherent control of the nuclear excitation with vacuum ultraviolet (VUV) lasers possible \cite{LarsIC2017,Liao_Coherent_2012}. 
Moreover, strong coupling of the nuclear transition to the electronic shell in $\isotope[229]{Th}$ ions may allow usage of the outer electrons to excite the nuclear isomeric state with optical lasers in a two-photon electronic bridge process~\cite{PorsevFlambaum_Brige_PRL_2010}.

Precise knowledge of the transition parameters such as energy and $\gamma$-decay rate (which in the theoretical treatment can be traced back to the reduced nuclear transition probabilities) is needed for the aforementioned applications. The first direct observation of decay of the low-energy isomeric state of $^{229}$Th  via internal
conversion (IC) has been reported  recently \cite{Wense_Nature_2016}. IC is very efficient in neutral Th atoms, as  the nuclear isomeric state $\isotope[229\rm{m}]{Th}$ decays via IC with enhancement factor $10^9$ with respect to the $\gamma$-decay~\cite{Wense_Nature_2016}. The most accepted value of the $\isotope[229\rm{m}]{Th}$ energy  $E_{\mathrm{m}}=7.8 \pm 0.5$~eV~\cite{Beck_78eV_2007,Beck_78eV_2007_corrected} could be determined only indirectly in a calorimetric measurement by
subtraction of x-ray energy differences between neighboring nuclear levels.  There are a few reasons today to doubt the precision of this value. In the first place, the  extraction of the energy value $E_{\mathrm{m}}$ from the experimental data in Refs.~\cite{Beck_78eV_2007,Beck_78eV_2007_corrected} was based on uncertain nuclear branching ratios, making possible an $E_{\mathrm{m}}$  value  outside  the error bars $\pm 0.5$~eV~\cite{Tkalya_PRC_2015}. Secondly, recent negative experimental results of two broadband photoexcitation attempts of the isomeric state may indicate that the transition energy lies in a different energy range \cite{Jeet_PRL_2015,Yamaguchi2015}. Furthermore, the recent results in Ref.~\cite{Seiferle_PRL_2017} on the short lifetime of the isomer in $\isotope{Th}^+$  may be indirect evidence that the IC channel is already open and  $E_{\mathrm{m}}$ is higher than the ionization potential of $\isotope{Th}^+$, i.e. approx.~12~eV~\cite{HerreraSancho_PRA_2012}. Otherwise only IC from excited electronic states is energetically allowed in Th ions ~\cite{Bilous_PRA_2017}.

In this work we propose an alternative to  measure the energy of the $\isotope[229]{Th}$ nuclear isomeric state at a precision typical to laser atomic spectroscopy, and to improve knowledge of its radiative lifetime. The method is based on nuclear deexcitation by laser-induced electronic bridge (LIEB). Especially for a nuclear transition energy $E_{\mathrm{m}}$ around 12 eV, this process has an especially high rate and could be very efficient to characterize the nuclear transition with the help of a tunable optical laser. The usual electronic bridge (EB) is a process coupling the nucleus to the atomic shell~\cite{Krutov_JETPLett_1990, StrizhovTkalya_JETP_1991,Karpeshin_PRL_1999,Kalman_PRC_2001}. It occurs when the nuclear transition energy is not sufficient for IC, but is close to an atomic transition energy. In order to fulfill energy conservation, the 
transfer of the nuclear excitation to a bound electron which undergoes a transition to an excited state is accompanied by the emission or absorption of a photon. The EB process might play a significant role in the decay of the isomeric state in Th ions, where the IC channel is energetically closed. 
Calculations have shown that the EB process can significantly change the isomeric state lifetime if the energy $E_{\mathrm{m}}$ happens to be close to  $M1$ transitions of the electronic shells ~\cite{PorsevFlambaum_Brige3+_PRA_2010,PorsevFlambaum_Brige1+_PRA_2010}.

LIEB is a version of EB process in which the additional photon is not emitted, but absorbed by the electronic shell from an externally applied laser field.  The process is thus equivalent to the excitation of the electronic shell with two photons, one of which is provided by the laser source and the other one from the nuclear deexcitation. This process was first considered for atomic $\isotope[229]{Th}$ in 1992 by F.~Karpeshin \textit{et al.}~\cite{Karpeshin_LIEB_1992}. It was shown that an enhancement factor of the decay of the isomeric state $\isotope[229\rm{m}]{Th}$ of the order $10^3$ can be achieved. We note that at the time, the isomeric energy was believed to lie around 3.5 eV, such that IC of the neutral thorium atom wouldn't have been possible. Meanwhile it is  known today that the isomeric nuclear state in a neutral atom decays very fast via IC \cite{Wense_Nature_2016}, making the proposed scheme obsolete. We consider LIEB in the  $\isotope[229]{Th}^{3+}$ ion and show that with up-to-date laser technology the enhancement can reach the order $10^8$ if the value $E_{\mathrm{m}}$ is close to the ionization threshold of $\isotope{Th}^+$. This would allow for a measurement of the transition energy and decay rate by using a tunable optical laser. $\isotope[229]{Th}^{3+}$ ions allow high controllability as far as trapping and cooling are concerned and possess a simple electronic spectrum, rendering a reliable experimental implementation possible.

\begin{figure}[ht!]
\centering
\includegraphics[width=0.6\textwidth]{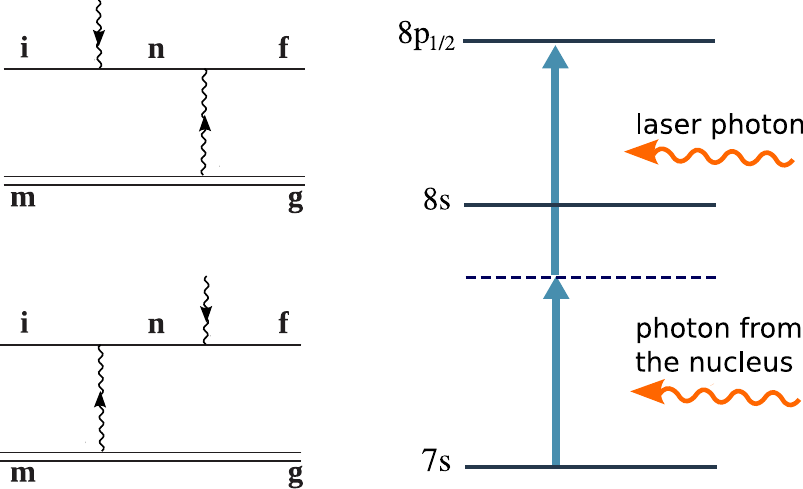}
\caption{Example of LIEB process (right)  and its Feynman diagrams (left).  The solid lines in the Feynman diagrams correspond to the electronic states, the double-lines represent the nuclear state and the wiggly lines depict the photons. See text for further explanations.}
\label{scheme}
\end{figure}

The LIEB process can be represented by two Feynman diagrams shown on the right-hand side of Fig.~\ref{scheme}. The states $\ket{m}$ and $\ket{g}$ denote the isomeric and the ground states of the nucleus, respectively. The states $\ket{i}$ and $\ket{f}$ in turn are the initial and the final electronic states, respectively. The intermediate electronic state $\ket{n}$ runs over all the levels allowed by the selection rules (in the whole spectrum, including the continuum). The transition $\isotope[229\rm{m}]{Th}\rightarrow\isotope[229\rm{g}]{Th}$ has $M1$ multipolarity. We make use of the fact  (also considered in Ref.~\cite{Karpeshin_LIEB_1992}) that the coupling of the nuclear transition is maximal to an $M1$ electronic transition between $s$ orbitals and considerable for transitions between $p_{1/2}$ orbitals. We consider therefore the excited $7s$ state  as the initial electronic state $\ket{i}$ and the $8p_j$ and $9p_j$  states with the total angular momentum $j\in\{1/2,3/2\}$  in the role of the state $\ket{f}$. With this choice  the channels $7s \rightarrow ns \rightarrow \ket{f}$ with $n=8,\, 9$  and 
$7s \rightarrow 7p_{1/2} \rightarrow \ket{f}$ would then play  the dominating role.
These two channels are described by the two different Feynman diagrams in Fig.~\ref{scheme}, corresponding to  different absorption sequences  of the two photons by the electron.   A further advantage of using the  $7s$ state as the initial state is its metastability, i.e. very long lifetime of~0.6~s~\cite{Safronova_ThIon_2013}. Note that since $\ket{n}$ is actually a virtual state,  the energy of the $M1$ photon does not necessarily have to coincide with the energy of the $M1$ electronic transition, though the overall transition energy $\ket{i}\rightarrow\ket{f}$ has to be equal to the sum of the energies of the two photons (see Fig.~\ref{scheme}).

The nuclear isomer energy $E_{\mathrm{m}}$ can be found via scanning with a tunable laser for a LIEB resonance, i.e., for the population and decay of the upper state $\ket{f}$. The decay observation of  the $\ket{f}$ state would correspond to  the sum of the nuclear excitation energy and the energy of the laser photon equaling the overall electronic transition energy. If the resonant laser photon frequency is $\omega$, then $E_{\mathrm{m}}$ can be calculated as
\begin{equation}
E_{\mathrm{m}} = E_f-E_i-\hbar \omega,
\end{equation}
where $E_i$ and $E_f$ are the energies of the electronic state $\ket{i}$ and $\ket{f}$ respectively and $\hbar$ is Planck constant. The height of the measured LIEB resonances would provide information about the magnitude of the nuclear $\gamma$-decay rate.

\section{Theory}

We calculate the LIEB rate  $W^\mathrm{LIEB}$ by relating  to  the rate of the inverse spontaneous process $W^\mathrm{spont}$ via the expression~\cite{Sobelman_book_1979}
\begin{equation}
W^\mathrm{LIEB}=W^\mathrm{spont} \frac{4\pi^3 c^2}{\hbar \omega^3} P_\omega \delta,
\end{equation}
where $\omega$ and $P_\omega$ are the frequency and the spectral intensity of the laser field, $c$ is the speed of light and
\begin{equation}
\delta=\frac{(2I_g+1)(2J_f+1)}{(2I_{\mathrm{m}}+1)(2J_i+1)}
\end{equation}
denotes the ratio of  magnetic quantum number degeneracies  of the initial states in the case of the direct and the inverse processes. These states are characterized by the nuclear spin $I$ and the total angular momentum of the electronic shell $J$. For calculation of $W^\mathrm{spont}$ we adopt the formalism developed in~\cite{PorsevFlambaum_Brige3+_PRA_2010} for the traditional EB scenario, taking into account the different direction of the nuclear process, i.e. excitation instead of deexcitation. The Feynman diagrams of this process can be obtained from the diagrams in Fig.~\ref{scheme} by mirror reflection and inversion of the photon lines. Based on the corresponding amplitudes, one can derive the rate of the decay $W^\mathrm{spont}$, which after averaging over the initial states, summation over the final states and application of Wigner-Eckart theorem \cite{Edmonds} can be rewritten with the help of the isomer $\gamma$-decay rate $\Gamma_\gamma$ as
\begin{equation}
W^\mathrm{spont}=\left(\frac{\hbar \omega}{E_{\mathrm{m}}}\right)^3\frac{G_1+G_{12}+G_2}{3(2J_f+1)} \cdot\frac{2I_{\mathrm{m}}+1}{2I_g+1} \cdot \Gamma_\gamma.
\end{equation}
Here, the ratio $\frac{2I_{\mathrm{m}}+1}{2I_g+1}$ takes into account different initial nuclear states for the considered spontaneous process and $\gamma$-decay. The quantities $G_i$ with $i=1,2$ corresponding to the two Feynman diagrams and their interference can be written with the help of the reduced matrix elements of the electric $D$ and magnetic $T$ dipole operators  \cite{Johnson_book_2007} of the valence electron giving the following expressions (in atomic units)
\begin{eqnarray}
G_1&=&\sum_{J_n}\frac{1}{2J_n+1}\nonumber\\
 &\times& \left| \sum_{\gamma_k} 
 \frac{\rbra{\gamma_i J_i}D\rket{\gamma_k J_n} \rbra{\gamma_k J_n} T \rket{\gamma_f J_f}}
 {E_f-E_k-E_{\mathrm{m}}}
 \right|^2
,\end{eqnarray}
\begin{eqnarray}
G_{12}&=& 2\sum_{J_t J_n} (-1)^{J_t+J_n}
\left\{
\begin{array}{ccc}
J_f & J_t & 1 \\
J_i & J_n & 1
\end{array}
\right\}
\nonumber\\
 &\times & \sum_{\gamma_k} 
 \frac{\rbra{\gamma_i J_i}D\rket{\gamma_k J_n} \rbra{\gamma_k J_n} T \rket{\gamma_f J_f}}
 {E_f-E_k-E_{\mathrm{m}}}\nonumber\\
 &\times& \sum_{\gamma_s} 
 \frac{\rbra{\gamma_i J_i}T\rket{\gamma_s J_t} \rbra{\gamma_s J_t} D \rket{\gamma_f J_f}}
 {E_i-E_s+E_{\mathrm{m}}}
,\end{eqnarray}
\begin{eqnarray}
G_2&=&\sum_{J_n}\frac{1}{2J_n+1}\nonumber\\
 &\times& \left| \sum_{\gamma_k} 
 \frac{\rbra{\gamma_i J_i}T\rket{\gamma_k J_n} \rbra{\gamma_k J_n} D \rket{\gamma_f J_f}}
 {E_i-E_k+E_{\mathrm{m}}}
 \right|^2
.\end{eqnarray}
The sums are carried out over the total angular momenta of the intermediate states $J_n$ and $J_t$ and over all other electronic quantum numbers denoted by the generic indices $\gamma_k$ and $\gamma_s$. The notation $\rbra{}\cdot\rket{}$  stands for the reduced matrix elements after application of the Wigner-Eckart theorem \cite{Edmonds}.

\section{Numerical Results}

For numerical evaluation of $W_\mathrm{LIEB}$ we require the nuclear $\gamma$-decay rate, the spectrum of the valence electron and the $E1$ and $M1$ matrix elements between the states $\ket{i}$ and $\ket{n}$ and the states $\ket{f}$ and $\ket{n}$ for all $\ket{n}$ allowed by the selection rules (including the continuum states). We consider the value $P_\omega=1 \; \frac{\mathrm{W}}{\mathrm{m}^2\cdot\mathrm{Hz}}$ for the optical laser spectral intensity. This can be achieved with modern tunable lasers in the range from 560 to 1000 nm. We assume $\Gamma_\gamma=10^{-4} \; \mathrm{1/s}$ corresponding to the recent theoretical value for the reduced nuclear matrix element $B_\downarrow=0.0076\;\mathrm{W.u.}$~\cite{Minkov_Palffy_PRL_2017}. This value is smaller than previously estimated \cite{Dyk98,Ruch2006} and provides therefore a more pessimistic result for the isomer  radiative decay rate. 

The spectrum of the valence electron is taken from the database in Ref.~\cite{DBlevels}. The electronic matrix elements are calculated based on \textit{ab initio} wave-functions evaluated employing the following procedure: First we calculate a few low-lying states of the valence electron using Dirac-Hartree-Fock method in the frozen-core approximation. These serve as initial estimates for more accurate wave functions calculated by considering the ion to be placed into a cavity of the radius $R=60\;\mathrm{a.u.}$ and building virtual orbitals 1--20$s$, 2--20$p$, 3--20$d$, 4--25$f$, 5--18$g$ via the expansion in a $B$-spline basis~\cite{Johnson_book_2007}. Due to the large size of the cavity, this procedure  does not affect the matrix elements considerably, but allows us to work with a discrete spectrum at positive energies. In order to benchmark our method we have constructed the same basis set as in Ref.~\cite{PorsevFlambaum_Brige3+_PRA_2010} and have reproduced numerical results presented therein. Our results show  good agreement with Ref.~\cite{PorsevFlambaum_Brige3+_PRA_2010}
confirming the reliability of the computed electronic wave functions and of our numerical results.

\begin{figure*}[ht!]
\centering
\includegraphics[width=1\textwidth]{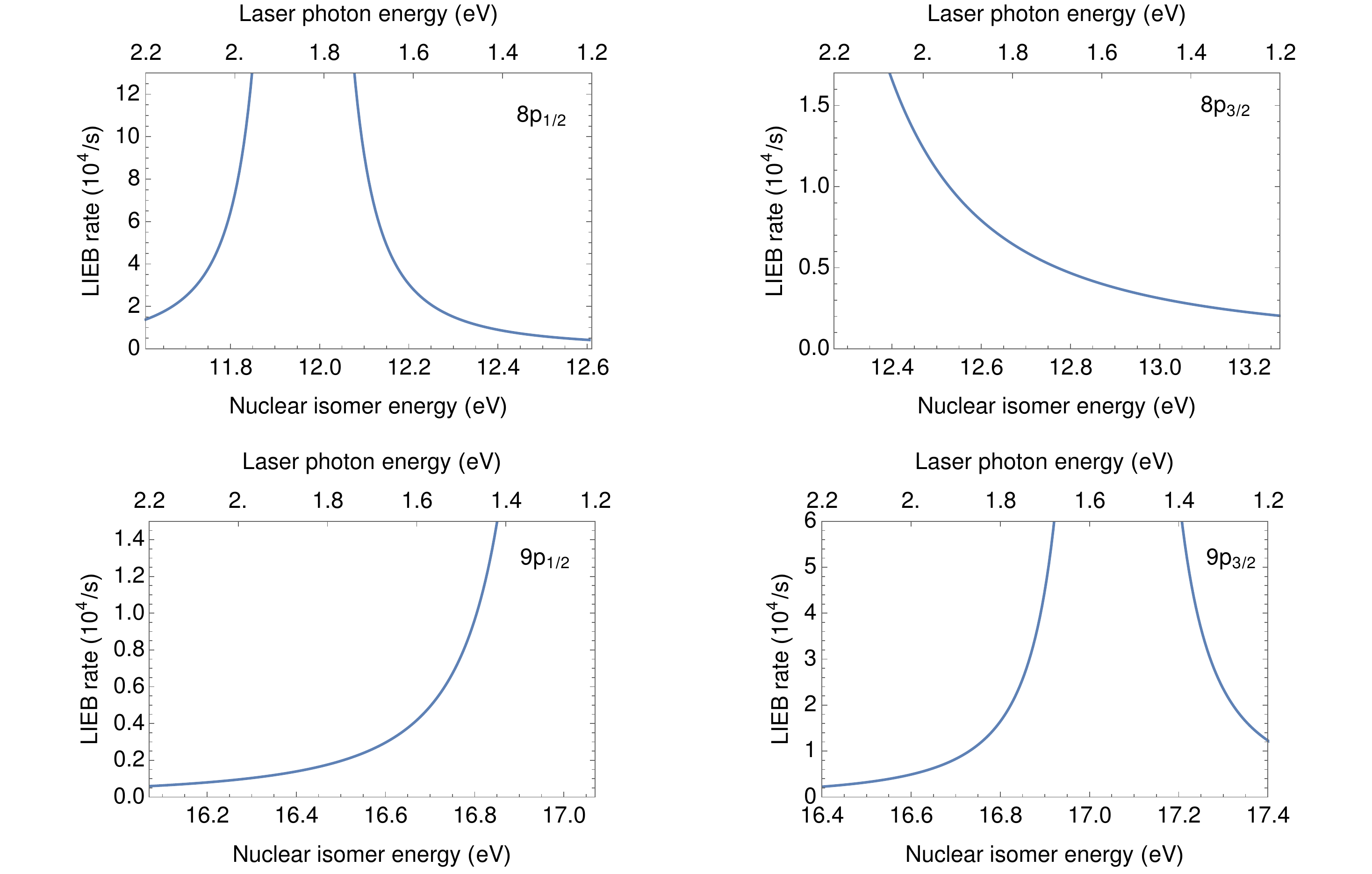}
 \caption{The LIEB rate as  a function of the nuclear isomeric state energy (lower axis) and the corresponding laser photon energy (upper axis). The final electronic states are $8p_{1/2}$, $8p_{3/2}$, $9p_{1/2}$ and $9p_{3/2}$, respectively. The considered laser spectral intensity  is $P_\omega=1 \; \frac{\mathrm{W}}{\mathrm{m}^2\cdot\mathrm{Hz}}$.}
\label{rate_opt}
\end{figure*}

\begin{figure*}[ht!]
\centering
\includegraphics[width=1\textwidth]{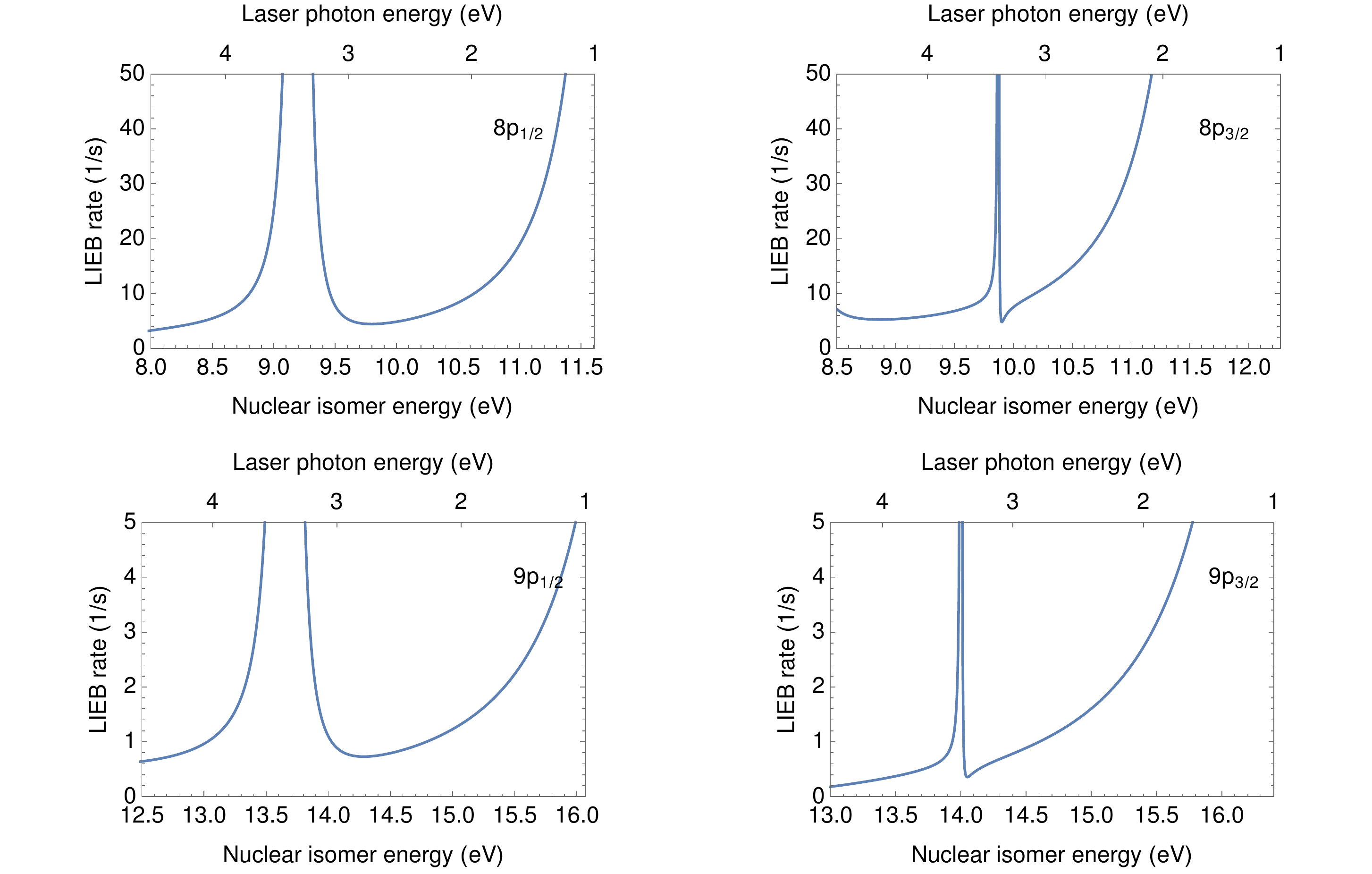}
 \caption{The LIEB rate as  a function of the nuclear isomeric state energy (lower axis) and the corresponding laser photon energy (upper axis). The final electronic states are $8p_{1/2}$, $8p_{3/2}$, $9p_{1/2}$ and $9p_{3/2}$,  respectively. The considered  laser spectral intensity is $P_\omega=10^{-2} \; \frac{\mathrm{W}}{\mathrm{m}^2\cdot\mathrm{Hz}}$.}
\label{rate_uv}
\end{figure*}

The graphs in Fig.~\ref{rate_opt} show the LIEB rate as  a function of the nuclear isomer energy for different final electronic states. We depict also the energy of the resonant laser photon $\hbar \omega$, considering  values in the range from 1.2 to 2.2~eV available with modern high power tunable lasers. If the energy $E_{\mathrm{m}}$ is close to the electronic $M1$ transition, then a large enhancement $W_\mathrm{LIEB}/\Gamma_\gamma$ up to $10^8$ can be achieved. The large peak appearing in all the four cases corresponds to the resonance in the lower diagram in Fig.~\ref{scheme} for the $M1$ electronic transition between the initial state $\ket{i}=\ket{7s}$ and the intermediate state $\ket{n}=\ket{8s}$ or $\ket{n}=\ket{9s}$.  We note however that this substantial enhancement would only be present if the isomer energy lies around 12 or 17 eV. 

In case $E_{\mathrm{m}}$ is closer to the presently accepted value of $7.8 \pm 0.5$~eV \cite{Beck_78eV_2007,Beck_78eV_2007_corrected}, a laser with higher photon energy is required.  
 In Fig.~\ref{rate_uv} we show the LIEB rate calculated using the value $P_\omega=10^{-2} \; \frac{\mathrm{W}}{\mathrm{m}^2\cdot\mathrm{Hz}}$ for  laser photon energies higher than 2.2~eV. 
This might be a too conservative estimate for 2.5 eV, but simultaneously already demanding for 4.5 eV photon energies, respectively. We observe new peaks, which correspond to  the case of resonances in the upper Feynman diagram in Fig.~\ref{scheme}, when $E_{\mathrm{m}}$ coincides with the energy between the final state and an intermediate state among $7p_j$ and $8p_j$ with $j\in\{1/2,3/2\}$.  In this case the strongest  peaks occur for  electronic $M1$ transitions between $p_{1/2}$ states.  Provided the spectral intensity $P_\omega=10^{-2} \; \frac{\mathrm{W}}{\mathrm{m}^2\cdot\mathrm{Hz}}$ in the depicted region of laser photon energy can be reached,  the described method would be applicable for lower values of $E_\mathrm{m}$ as low as 8~eV under  conditions of  good detection efficiency.

\section{Discussion}

Using the described scheme the energy of the VUV nuclear transition $\isotope[229\rm{m}]{Th}\rightarrow\isotope[229\rm{g}]{Th}$ can be determined using a tunable optical laser. The rate of the LIEB process can be measured by means of detection of the spontaneous decay photons  of the final electronic state $\ket{f}$. Comparing the value of the resonant LIEB rate to the theoretical predictions, it is possible to find $\Gamma_\gamma$. Note that among various charge states of thorium, the monovalent  $\isotope[229]{Th}^{3+}$ ion allows particularly accurate  \textit{ab initio} calculations for the electronic shell. The $\isotope[229]{Th}^{2+}$ ions which would also fulfill this requirement  cannot be used for the described method, as  no metastable states with either $s$- or $p_{1/2}$-electrons involved are available. The described scheme in  $\isotope[229]{Th}^{3+}$ is therefore a unique reliable way for determination of the Th nuclear isomeric state parameters via laser-induced nuclear deexcitation involving the electronic shell.

The LIEB scheme requires production and trapping of $\isotope[229\rm{m}]{Th}$ in the $3+$ ionization state and the excitation of the electronic shell to the $7s$ initial state. Regarding the production of the isomeric state, the direct nuclear photoexcitation of the ground-state thorium has failed so far due to the poor knowledge of the transition energy. Instead, one typically takes advantage of the 2\% branching ratio to the low-energy isomeric state \cite{twoperc} in the process of  $\alpha$-decay of $\isotope[233]{U}$ and subsequent gamma-ray de-excitation. Inevitably a mixture of mostly ground state and just few isomeric state $\isotope[229]{Th}^{3+}$ ions would then be loaded in the ion trap. We anticipate that $10^4-10^5$ ions could be produced and trapped at the Munich setup~\cite{Wense_Nature_2016,Seiferle_PRL_2017} where a cryogenic Paul trap is being built~\cite{Schwarz_RSInstr_2012}. This  setup will keep $\isotope[229]{Th}^{3+}$ ions in the absence of charge exchange and chemical reactions for a storage time longer than the time of isomer radiative decay. The electronic shell can be effectively transferred from the ground state $5f_{5/2}$ to the $7s$ state making use of the Stimulated Raman Adiabatic Passage (STIRAP) method~\cite{Schore_PRA_1992} via the intermediate state $6d_{3/2}$. Note that the second transition in the STIRAP scheme is an electric quadrupole ($E2$) transition and thus requires higher driving laser intensity. Once reached, the $7s$ state has a long lifetime (0.6 s) allowing for applying the LIEB scheme.

Upon interaction with the tunable optical or ultra-violet laser, the isomeric state which otherwise has a long radiative lifetime of approx. $10^4$~s may decay via the LIEB process. Population and radiative decay of the upper electronic states $np_j$ with $n=8,\, 9$ and $j\in\{1/2,3/2\}$ could be observed by detection of  fluorescence photons. We present the calculated radiative decay rates for the electric dipole channels together with experimental photon energy values from Ref.~\cite{DBlevels} in  Tables~\ref{table_rdecay1}---\ref{table_rdecay4}.
Since the energy of the emitted photon in the dominant decay channels is mostly in the range 15--20 eV, i.e. significantly higher than that of the laser photons that are used for the excitation, the selection of a suitable photocathode material (like CsI) for photoelectron production inside the vacuum system will ensure a detection that is free from  background of laser stray light. The background due to the absorption of the laser radiation by the electron in thorium ions with the nucleus in the ground state is expected to be small since the envisaged electronic transitions are energetically much higher than the laser photon energy. A direct comparison with the case of $\isotope[232]{Th}$ which does not possess any nuclear states at optical energies should help to identify the signal stemming from the isomeric state decay.

\begin{table}[h!]
\centering
\begin{tabular}{|c|c|c|}
\hline
Final state & Photon energy (eV) & Rate  ($10^8/\rm{s}$) \\
\hline
 $7s_{1/2}$ & 13.8   &     0.26 \\
 \hline
 $8s_{1/2}$&   1.85   &     4.96\\
 \hline
$6d_{3/2}$ & 15.5    &    19.4\\
 \hline
 $7d_{3/2}$ &  1.84    &    5.81\\
 \hline
\end{tabular}
\caption{Radiative decay channels of the  $8p_{1/2}$ state.}\label{table_rdecay1}
\end{table}

\begin{table}[h!]
\centering
\begin{tabular}{|c|c|c|}
\hline
 Final state & Photon energy (eV) & Rate  ($\frac{10^8}{\rm{sec}}$)\\
\hline
 $7s_{1/2}$ &  14.5        &5.05\\
  \hline
  $8s_{1/2}$ &  2.51        &11.3\\
  \hline
$6d_{3/2}$  & 16.2    &    3.12\\
  \hline
$6d_{5/2}$  & 15.6  &      24.9\\
  \hline
$7d_{3/2}$ &  2.50     &   1.15\\
\hline
$7d_{5/2}$ &  2.29    &    8.76\\
  \hline
\end{tabular}
\caption{Radiative decay channels of the $8p_{3/2}$ state.}\label{table_rdecay2}
\end{table}

\begin{table}[h!]
\centering
\begin{tabular}{|c|c|c|}
\hline
Final state & Photon energy (eV) & Rate  ($\frac{10^8}{\rm{sec}}$) \\
\hline
 $7s_{1/2}$ &18.3 & 0.57\\
 \hline
 $8s_{1/2}$&   6.31& 0.045 \\
 \hline
  $9s_{1/2}$& 1.21& 3.30 \\
 \hline
 $6d_{3/2}$ & 20.0& 11.5 \\
 \hline
 $7d_{3/2}$ & 6.30& 1.36\\
 \hline
  $8d_{3/2}$ &   1.06& 3.13 \\
 \hline
\end{tabular}
\caption{Radiative decay channels of the $9p_{1/2}$ state.}\label{table_rdecay3}
\end{table}

\begin{table}[h!]
\centering
\begin{tabular}{|c|c|c|}
\hline
Final state & Photon energy (eV) & Rate  ($\frac{10^8}{\rm{sec}}$)\\
\hline
 $7s_{1/2}$ & 18.6& 1.76 \\
  \hline
  $8s_{1/2}$ &  6.64& 1.30\\
  \hline
  $9s_{1/2}$ & 1.54& 6.01\\
  \hline
$6d_{3/2}$  &  20.3& 1.80 \\
  \hline
$6d_{5/2}$  & 19.68 & 14.5\\
  \hline
$7d_{3/2}$ &  6.63& 0.42 \\
\hline
$7d_{5/2}$ &   6.42& 3.02 \\
\hline
$8d_{3/2}$ &  1.39 & 0.56 \\
\hline
$8d_{5/2}$ & 1.32& 4.73\\
  \hline
\end{tabular}\label{table_rdecay4}
\caption{Radiative decay channels of the $9p_{3/2}$ state.}\label{table_rdecay4}
\end{table}


\begin{ack}
The authors gratefully acknowledge funding by the EU FET-Open project 664732. 
\end{ack}

\section*{References}
\bibliography{refs}

\providecommand{\newblock}{}
\begin{thebibliography}{10}
\expandafter\ifx\csname url\endcsname\relax
  \def\url#1{{\tt #1}}\fi
\expandafter\ifx\csname urlprefix\endcsname\relax\def\urlprefix{URL }\fi
\providecommand{\eprint}[2][]{\url{#2}}

\bibitem{Wense_Nature_2016}
von~der Wense L, Seiferle B, Laatiaoui M, Neumayr J~B, Maier H~J, Wirth H~F,
  Mokry C, Runke J, Eberhardt K, D{\"u}llmann C~E, Trautmann N~G and Thirolf
  P~G 2016 {\em Nature\/} {\bf 533} 47--51 ISSN 0028-0836 article
  \urlprefix\url{http://dx.doi.org/10.1038/nature17669}

\bibitem{Beck_78eV_2007}
Beck B~R, Becker J~A, Beiersdorfer P, Brown G~V, Moody K~J, Wilhelmy J~B,
  Porter F~S, Kilbourne C~A and Kelley R~L 2007 {\em Phys. Rev. Lett.\/} {\bf
  98}(14) 142501
  \urlprefix\url{https://link.aps.org/doi/10.1103/PhysRevLett.98.142501}

\bibitem{Beck_78eV_2007_corrected}
 2009 {\em Improved Value for the Energy Splitting of the Ground-State Doublet
  in the Nucleus $^{229m}\mathrm{Th}$\/} vol LLNL-PROC-415170

\bibitem{Peik_Clock_2003}
{E Peik} and {Chr Tamm} 2003 {\em Europhys. Lett.\/} {\bf 61} 181--186
  \urlprefix\url{https://doi.org/10.1209/epl/i2003-00210-x}

\bibitem{Campbell_Clock_2012}
Campbell C~J, Radnaev A~G, Kuzmich A, Dzuba V~A, Flambaum V~V and Derevianko A
  2012 {\em Phys. Rev. Lett.\/} {\bf 108}(12) 120802
  \urlprefix\url{https://link.aps.org/doi/10.1103/PhysRevLett.108.120802}

\bibitem{Peik_Clock_2015}
Peik E and Okhapkin M 2015 {\em Comptes Rendus Physique\/} {\bf 16} 516 -- 523
  ISSN 1631-0705 the measurement of time / La mesure du temps
  \urlprefix\url{http://www.sciencedirect.com/science/article/pii/S1631070515000213}

\bibitem{Tkalya_NuclLaser_2011}
Tkalya E~V 2011 {\em Phys. Rev. Lett.\/} {\bf 106}(16) 162501
  \urlprefix\url{https://link.aps.org/doi/10.1103/PhysRevLett.106.162501}

\bibitem{LarsIC2017}
von~der Wense L, Seiferle B, Stellmer S, Weitenberg J, Kazakov G, P\'alffy A
  and Thirolf P~G 2017 {\em Phys. Rev. Lett.\/} {\bf 119} in press

\bibitem{Liao_Coherent_2012}
Liao W~T, Das S, Keitel C~H and P\'alffy A 2012 {\em Phys. Rev. Lett.\/} {\bf
  109}(26) 262502
  \urlprefix\url{https://link.aps.org/doi/10.1103/PhysRevLett.109.262502}

\bibitem{PorsevFlambaum_Brige_PRL_2010}
Porsev S~G, Flambaum V~V, Peik E and Tamm C 2010 {\em Phys. Rev. Lett.\/} {\bf
  105}(18) 182501
  \urlprefix\url{https://link.aps.org/doi/10.1103/PhysRevLett.105.182501}

\bibitem{Tkalya_PRC_2015}
Tkalya E~V, Schneider C, Jeet J and Hudson E~R 2015 {\em Phys. Rev. C\/} {\bf
  92}(5) 054324
  \urlprefix\url{https://link.aps.org/doi/10.1103/PhysRevC.92.054324}

\bibitem{Jeet_PRL_2015}
Jeet J, Schneider C, Sullivan S~T, Rellergert W~G, Mirzadeh S, Cassanho A,
  Jenssen H~P, Tkalya E~V and Hudson E~R 2015 {\em Phys. Rev. Lett.\/} {\bf
  114}(25) 253001
  \urlprefix\url{https://link.aps.org/doi/10.1103/PhysRevLett.114.253001}

\bibitem{Yamaguchi2015}
Yamaguchi A, Kolbe M, Kaser H, Reichel T, Gottwald A and Peik E 2015 {\em New
  Journal of Physics\/} {\bf 17} 053053
  \urlprefix\url{http://stacks.iop.org/1367-2630/17/i=5/a=053053}

\bibitem{Seiferle_PRL_2017}
Seiferle B, von~der Wense L and Thirolf P~G 2017 {\em Phys. Rev. Lett.\/} {\bf
  118}(4) 042501
  \urlprefix\url{https://link.aps.org/doi/10.1103/PhysRevLett.118.042501}

\bibitem{HerreraSancho_PRA_2012}
Herrera-Sancho O~A, Okhapkin M~V, Zimmermann K, Tamm C, Peik E, Taichenachev
  A~V, Yudin V~I and G\l{}owacki P 2012 {\em Phys. Rev. A\/} {\bf 85}(3) 033402
  \urlprefix\url{https://link.aps.org/doi/10.1103/PhysRevA.85.033402}

\bibitem{Bilous_PRA_2017}
Bilous P~V, Kazakov G~A, Moore I~D, Schumm T and P\'alffy A 2017 {\em Phys.
  Rev. A\/} {\bf 95}(3) 032503
  \urlprefix\url{https://link.aps.org/doi/10.1103/PhysRevA.95.032503}

\bibitem{Krutov_JETPLett_1990}
Krutov V 1990 {\em JETP Lett.\/} {\bf 52} 584--588

\bibitem{StrizhovTkalya_JETP_1991}
Strizhov V and Tkalya E 1991 {\em Sov. Phys. JETP\/} {\bf 72} 387

\bibitem{Karpeshin_PRL_1999}
Karpeshin F~F, Band I~M, Trzhaskovskaya M~B and Pastor A 1999 {\em Phys. Rev.
  Lett.\/} {\bf 83}(5) 1072--1072
  \urlprefix\url{https://link.aps.org/doi/10.1103/PhysRevLett.83.1072}

\bibitem{Kalman_PRC_2001}
K\'alm\'an P and B\"ukki T 2001 {\em Phys. Rev. C\/} {\bf 63}(2) 027601
  \urlprefix\url{https://link.aps.org/doi/10.1103/PhysRevC.63.027601}

\bibitem{PorsevFlambaum_Brige3+_PRA_2010}
Porsev S~G and Flambaum V~V 2010 {\em Phys. Rev. A\/} {\bf 81}(3) 032504
  \urlprefix\url{https://link.aps.org/doi/10.1103/PhysRevA.81.032504}

\bibitem{PorsevFlambaum_Brige1+_PRA_2010}
Porsev S~G and Flambaum V~V 2010 {\em Phys. Rev. A\/} {\bf 81}(4) 042516
  \urlprefix\url{https://link.aps.org/doi/10.1103/PhysRevA.81.042516}

\bibitem{Karpeshin_LIEB_1992}
Karpeshin F, Band I, Trzhaskowskaya M and Zon B 1992 {\em Physics Letters B\/}
  {\bf 282} 267 -- 270 ISSN 0370-2693
  \urlprefix\url{http://www.sciencedirect.com/science/article/pii/037026939290636I}

\bibitem{Safronova_ThIon_2013}
Safronova M~S and Safronova U~I 2013 {\em Phys. Rev. A\/} {\bf 87}(6) 062509
  \urlprefix\url{https://link.aps.org/doi/10.1103/PhysRevA.87.062509}

\bibitem{Sobelman_book_1979}
Sobelman I~I 1979 {\em Atomic Spectra And Radiative Transitions\/}
  (Springer-Verlag, Berlin)

\bibitem{Edmonds}
Edmonds A~R 1996 {\em Angular Momentum in Quantum Mechanics\/} (Princeton
  University Press, Princeton)

\bibitem{Johnson_book_2007}
Johnson W~R 2007 {\em Atomic Structure Theory: Lectures on Atomic Physics\/}
  (Springer, New York)

\bibitem{Minkov_Palffy_PRL_2017}
Minkov N and P\'alffy A 2017 {\em Phys. Rev. Lett.\/} {\bf 118} 212501

\bibitem{Dyk98}
Dykhne A~M and Tkalya E~V 1998 {\em JETP Lett.\/} {\bf 67} 251

\bibitem{Ruch2006}
Ruchowska E, P\l{}\'ociennik W~A, \ifmmode~\dot{Z}\else \.{Z}\fi{}ylicz J, Mach
  H, Kvasil J, Algora A, Amzal N, B\"ack T, Borge M~G, Boutami R, Butler P~A,
  Cederk\"all J, Cederwall B, Fogelberg B, Fraile L~M, Fynbo H~O~U, Hageb\o{}
  E, Hoff P, Gausemel H, Jungclaus A, Kaczarowski R, Kerek A, Kurcewicz W,
  Lagergren K, Nacher E, Rubio B, Syntfeld A, Tengblad O, Wasilewski A~A and
  Weissman L 2006 {\em Phys. Rev. C\/} {\bf 73}(4) 044326

\bibitem{DBlevels}
\url{http://web2.lac.u-psud.fr/lac/Database/Tab-energy/Thorium/} accessed:
  2017-05-23

\bibitem{twoperc}
Barci V, Ardisson G, Barci-Funel G, Weiss B, El~Samad O and Sheline R~K 2003
  {\em Phys. Rev. C\/} {\bf 68}(3) 034329

\bibitem{Schwarz_RSInstr_2012}
Schwarz M, Versolato O, Windberger A, Brunner F, Ballance T, Eberle S, Ullrich
  J, Schmidt P, Hansen A, Gingell A, Drewsen M and Crespo L\'opez-Urrutia J
  2012 {\em Review of Scientific Instruments\/} {\bf 83} 083115
  (\textit{Preprint} \eprint{http://dx.doi.org/10.1063/1.4742770})
  \urlprefix\url{http://dx.doi.org/10.1063/1.4742770}

\bibitem{Schore_PRA_1992}
Shore B~W, Bergmann K, Kuhn A, Schiemann S, Oreg J and Eberly J~H 1992 {\em
  Phys. Rev. A\/} {\bf 45}(7) 5297--5300
  \urlprefix\url{https://link.aps.org/doi/10.1103/PhysRevA.45.5297}

\end{thebibliography}

\end{document}